\documentstyle[12pt]{article}
\def\preprintno{IKDA 93/26 \\ hep-ph/9308262}
\makeatletter
\def\@maketitle{\newpage \null
  \@ifundefined{preprintno}
  {\vskip 2em}              % this is standard article.sty w/o preprint #
  {                         % \else
    \vspace*{-1\headsep}    % Move up to header region
    \vspace*{-1\headheight}
    \begin{flushright}      % insert Preprint number(s)
      \large \preprintno
    \end{flushright}
    \vspace*{-3em}
    \vskip \headsep	    % Move back down by part of what we moved up
    \vskip \headheight
    }                       % \fi for our modification
  \begin{center}            % from here on old definition
    {\LARGE \@title \par} \vskip 1.5em {\large \lineskip .5em
      \begin{tabular}[t]{c}\@author
      \end{tabular}\par}
    \vskip 1em {\large \@date} \end{center}
    \par
    \vskip 1.5em}
\makeatother
\def\slash#1{#1 \hskip -0.5em / }
\def\sint{\Sigma \hskip -1.0em \int}
\title{Operator Product Expansion  \\ for Inclusive Semileptonic Decays  \\
       in Heavy Quark Effective Field Theory}
\author{{\sc Thomas Mannel} \vspace*{5mm} \\
        Institut f\"ur Kernphysik \\
        Technische Hochschule Darmstadt \\
        Schlossgartenstr. 9, D--6100 Darmstadt \\
        Germany }
\begin{document}
\maketitle
\vfill
\begin{abstract}
\noindent
Inclusive semileptonic decays are discussed in the framework of
heavy quark effective field theory by employing the short distance
expansion in the effective theory. The lowest order term turns out
to be the parton model; the higher order terms may be regarded as
correction terms to the parton model result. The first nonvanishing
corrections to the parton model result are given and the lepton
energy spectrum of inclusive semileptonic decays of heavy mesons
is calculated.
\end{abstract}
\thispagestyle{empty}
\newpage
\setcounter{page}{1}
\section{Introduction}
Heavy quark effective field theory \cite{IW90,HQET} has turned out to be a
useful tool
for the description of heavy quark systems. Due to additional symmetries
\cite{IW90}
emerging in the heavy quark limit several predictions for the decays of
heavy hadrons may be obtained which are completely model independent.
In particular, the form factors of weak decays involving a heavy to
heavy transition are
severely constrained by heavy quark symmetry and some predictions
may be obtained even for heavy to light deacays \cite{h2l}.

The heavy quark expansion has also been applied to inclusive decays.
The first discussion of inclusive semileptonic decays in this framework
has been given by Chay, Georgi and Grinstein
in  \cite{CGG90}. Along similar lines proceeds the
work of Bigi, Shifman, Uraltsev and Vainstein \cite{BUV} and of  
Blok, Koyrakh, Shifman and Vainshtein \cite{BS93},
where inclusive
nonleptonic as well as semileptonic decays have been considered.

Usually the inclusive decay rates of heavy hadrons
and their spectra are approximated
by the decay of the heavy quarks in a parton model approximation. However,
this is a model approach and an estimate
of its error is difficult. This in is contrast to the methods described
in \cite{CGG90,BUV} and also to the present approch which is a
controllable approximation to QCD corresponding to expansions in small
parameters.

The method proposed in \cite{CGG90,BUV} to deal with inclusive processes
is to use an operator product expansion, which is justified for heavy
hadrons due to its large mass. The lowest dimension operator is a
dimension three operator; its matrix elements are normalized due to
heavy quark symmetries and yield the parton model result. Corrections
enter through operators of dimension five; their matrix elements
are parameters to be taken from measurements. These corrections have
been calculated for nonleptonic and for semileptonic decays
of mesons in \cite{BUV} and for the differential distributions in 
semileptonic decays in \cite{BS93}; 
semileptonic decays of baryons have been considered in \cite{MW93}.

In the present paper we shall consider only semileptonic decays
and formulate a slightly different approach as the one
proposed in \cite{BUV,BS93}. After switching to heavy quark effective
theory the space time dependence of the heavy quark
fields in the hadronic tensor
is only due to the residual momentum $k$ of the heavy quark $Q$
inside the heavy hadron $H$, i.e.~$k = P_H - m_Q v $.
In semileptonic decays there is in addition to the scale set by
the heavy mass $m_Q$ also the
momentum transfer to the leptons $q^2$.
After switching to the effective theory the relevant
variable becomes ${\cal Q}^2 = (q - m_Q v)^2$, which defines
in addition to the mass $m_Q$ of the heavy quark a second
scale in the problem.

In large portions of the phase space ${\cal Q}^2$ is of the order of
$m_Q^2$ and thus a simultaneous expansion in both $1/m_Q$ and
$1/{\cal Q}^2$ is appropriate. This is the approach chosen in
\cite{BUV,BS93}\footnote{I thank Mark Wise for a clarifying dicussion on
                    this point.}.
However, one may also consider the region in phase space, where
$$
m_Q^2 \gg (q-m_Q v)^2 \gg \Lambda_{QCD}^2
$$
and thus an expansion only in powers of $1/{\cal Q}^2$ is necessary
while only the leading term of the $1/m_Q$ is kept.

In the present paper we shall consider the latter approach and compare
to the one of \cite{BUV,BS93}.
The paper is organized as follows. In the next section we fix our
notation by giving the basic formulae for the inclusive
decay rates. In section 3 we set up the operator product expansion
in the effective theory and discuss the lowest order term and the
first nonvanishing correction. The setup given here may be used for
baryonic as well as for mesonic decays and in section 4 we give
a parametrization for the hadronic matrix elements for meson- and
baryon inclusive decays and discuss short distance QCD corrections.
Finally, in section 5 we discuss the lepton energy spectrum in
inclusive semileptonic decays in some detail.

\section{Inclusive Decay: Matrix Elements and Rates}
In order to set up the notation
we shall briefly recall the basic formulae for the
description of inclusive semileptonic decays. The inclusive
semileptonic decay of a heavy hadron $H$ (either a heavy 
meson or a heavy baryon)
into some hadronic state $X$ and two leptons
\begin{equation}
H(P_H = m_H v) \to X(P_X) + \ell (k) + \bar{\nu}_{\ell} (k')
\end{equation}
is mediated by an effective Hamiltonian of the form
\begin{equation}
H_{eff} = \frac{G_F}{\sqrt{2}} V_{Qq}
(\bar{Q} \gamma_\mu (1-\gamma_5) q )
(\bar{\ell} \gamma_\mu (1-\gamma_5) \nu_\ell)
\end{equation}
where $Q$ is the heavy quark contained in the in the heavy hadron
$H$\footnote{We shall assume that the flavor of $q$ may
             be determined from the hadronic state $X$.}.
The differential decay rate may be written in terms of the hadronic
tensor $W_{\mu \nu} (q,v)$
\begin{eqnarray}
W_{\mu \nu} (q,v) &=& \sint_X (2 \pi)^4 \delta^4 (P_H - q - P_X )
\\ \nonumber &&
< H(v) | (\bar{Q} \gamma_\mu (1-\gamma_5) q ) | X >
< X | (\bar{q} \gamma_\mu (1-\gamma_5) Q ) | H(v) > .
\end{eqnarray}
We shall consider only the case where the spins of the final state
leptons are not measured; hence the leptonic tensor is given by
\begin{equation}
\Lambda_{\mu \nu} = 8 \left( k_\mu k_\nu^\prime + k_\mu^\prime k_\nu
                    - g_{\mu \nu} (kk^\prime)
+ i \epsilon_{\mu \nu \alpha \beta} k^\alpha k^{\prime \beta} \right)
\end{equation}
The differential decay rate in the rest frame of the heavy hadron
is then 
\begin{equation}
d \Gamma = \frac{G_F^2}{4 m_H} | V_{Qq} |^2 W_{\mu \nu} (k+k^\prime,v)
\Lambda^{\mu \nu} d(PS)
\end{equation}
where $d(PS)$ is the phase space integral 
\begin{equation}
d(PS) = \int \widetilde{dk} \widetilde{dk}^\prime
        \delta (\mbox{observables})
        \quad
\widetilde{dk} = \frac{d^4 k}{(2 \pi)^3} \delta(k^2 - m_{Lep}^2)
                 \Theta (k_0) .
\end{equation}
The delta function $\delta$(observables) projects out the
observables to be considered; for the case of the energy spectrum
of the charged lepton in the rest frame of the decaying heavy meson
it is
$ \delta (\mbox{observables}) = \delta (E - vk^\prime) $,
since $k^\prime$ is the momentum of the charged lepton.
Furthermore, $m_{Lep}$ is the lepton mass which we shall neglect in the
following.

The hadronic tensor may be rewritten using standard techniques
\begin{equation}
W_{\mu \nu} (q,v) = \int d^4 x e^{iqx}
< H(v) | (\bar{Q}(x) \gamma_\mu (1-\gamma_5) q(x))
         (\bar{q}(0) \gamma_\nu (1-\gamma_5) Q(0))| H(v) >
\end{equation}
where we have explicitly spelled out the $x$ dependence of the
field operators.

The tensor $W_{\mu \nu}$ may be decomposed into scalar form
factors which then depend only on two invariants; these may be
cenveniently chosen to be $q^2$ and $vq$.
As discussed in some detail in \cite{CGG90}, one may relate these
form factors to the disconitinuity of the ones defining
the time ordered product
\begin{eqnarray} \label{Tprod}
T_{\mu \nu} (q,v) &=& \int d^4 x e^{iqx} \\ \nonumber
&& < H(v) | T \left[ (\bar{Q}(x) \gamma_\mu (1-\gamma_5) q(x))
         (\bar{q}(0) \gamma_\mu (1-\gamma_5) Q(0)) \right] | H(v) >
\end{eqnarray}
For a fixed value of $q^2$ there are two cuts of (\ref{Tprod})
in the complex $vq$ plane. A physical cut extents along the real
axis for
\begin{equation} \label{pcut}
\sqrt{q^2} < vq < \frac{1}{2m_H} (m_H^2 + q^2 - \mu^2)
\end{equation}
where $\mu$ is the mass of the lightest hadronic final state. In
addition, there is a cut extenting along the real axis for
\begin{equation} \label{upcut}
vq > \frac{1}{2m_H} ((2m_H+\mu)^2 - m_H^2 - q^2)
\end{equation}
The problematic region in phase space is the one where the invariant
mass of the hadronic final state approaches $\mu$; here resonances will
dominate and perturbative QCD will receive large corrections. However,
even in this region the predictions of perturbative QCD will be
recovered by a suitable smearing, i.e.~by calculating a smooth average along
some portion of the cut. This has been discussed in some detail 
for the case of $e^+ e^- \to $ hadrons in \cite{ave} and applied to the 
present case in \cite{CGG90}. In particular, 
one may expect a suitable smearing, if the energy spectrum of the 
charged lepton is calculated.

In order to calculate such an average one has to perform a contour 
integration of $T_{\mu \nu}$ in the complex $vq$ plane with a suitable 
weight function. The contour encircles the physical cut in such a
way that the discontinuity $T_{\mu \nu}$ (i.e.~the weighted hadronic 
tensor $W_{\mu \nu}$) of the cut is picked up.
As has been argued in \cite{CGG90} one may deform the
integration contours in the complex $vq$ plane 
in such a way that one always stays
away from the dangerous region by a distance large compared to
$\Lambda_{QCD}$.
This may be achived provided that
$\mu \gg \Lambda_{QCD}$ because for $\mu \to 0$ the cut (\ref{pcut})
and (\ref{upcut}) pinch the integration contour for
$q^2 = q_{max}^2 = (m_H - \mu)^2$. From these arguments one expects
that perturbative QCD may be applied after suitable smearing and for
$\mu \gg \Lambda_{QCD}$ which means that we may only study
inclusive semileptonic $b \to c$ transitions in perturbation theory,
while $b \to u$ transitions are likely to receive large, nonperturbative
corrections, at least in the region, where $q^2$ is close to its 
maximal value. 

Keeping this in mind one may relate
the perturbatively calculated time ordered product (\ref{Tprod})
to the hadronic tensor using the relation
\begin{equation}
W_{\mu \nu} (q,v) = 2 \mbox{ Im } T_{\mu \nu}
\end{equation}

The matrix element of the time ordered product in (\ref{Tprod}) 
may be expansed in a sum of local operators by performing a 
short distance expansion. However, there are still two large 
momenta involved, the momentum of the heavy meson which scales 
with $m_Q$ and the momentum transfer to the leptons $q^2$. 
In order to disentangle long and short distance contributions 
it is convenient to switch to the effective theory. In this 
process the heavy hadron momentum is rescaled and only the 
small residual momentum remains. As will become clear below, 
the relevant momentum transfer variable will be 
${\cal Q} = q-m_Q v$ which we assume to be the only large scale
in the problem. Once we are working in the effective theory.
an operator product expansion will be an expansion in inverse 
powers of ${\cal Q}^2$.

\section{Operator Product Expansion in the Effective Field Theory}
Switching to the effective field theory allows to scale out the
large parameter $m_H$. The relevant momentum variable characterizing
the initial state is then
a small momentum, namely the residual momentum $k = P_H - m_Q v$
of the heavy quark.
Matching the effective and the full theory to
lowest order in the $1/m_Q$ and $\alpha_s (m_Q)$ expansion one has
to replace
\begin{equation}
Q (x) \to e^{-im_Q v \cdot x} \frac{\slash{v} + 1}{2} h_v (x)
\end{equation}
where the $x$ dependence of the static heavy quark field $h_v$ is
due to the residual momentum $k$.

Injecting this into the hadronic tensor one obtains
\begin{eqnarray} \label{Htens}
W_{\mu \nu} (q,v) &=& \int d^4 x e^{ix(q-m_Q v)} \\ \nonumber
&& < H(v) | (\bar{h}_v(x) \gamma_\mu (1-\gamma_5) q(x))
         (\bar{q}(0) \gamma_\mu (1-\gamma_5) h_v (0))| H(v) >
\end{eqnarray}

An operator product expansion (i.e. a short distance expansion)
is justified, if the momentum
transfer variable ${\cal Q} = q-m_Q v$ is large compared to the residual
momentum $k$. In addition, switching to heavy quark effective theory
implies an expansion in powers of $1/m_Q$. In contrast to \cite{BUV,BS93}
we shall consider the portion of phase space in which
${\cal Q}^2 \ll m_Q^2$ and thus the lowest order term in the $1/m_Q$
expansion is sufficient. Hence we may use (\ref{Htens}) as the starting
point.

Using momentum conservation $P_H = m_Q v + k = q + P_X$ we find
that an operator product expansion is possible for $P_X^2 \gg m_X^2$
and that this expansion is a power series expansion in powers of
$\Lambda_{QCD} / m_X$, up to logarithmic terms which may be resummed
by renormalization group techniques. Note that, if $m_X$ is much
smaller than $m_H$, the use of the effective theory to lowest order
is justified, since higher order corrections to the effective theory
will be of the order $\Lambda_{QCD} / m_H$ and thus will be much
smaller.
 
However, as argued above, we shall only consider $b \to c$ inclusive 
transitions and thus $m_X \gg \Lambda_{QCD}$. For this case we may 
use the framework of perturbative QCD to calculate the coefficients
of the operator product expansion.

Thus we are lead to consider the
time ordered product of the two currents
\begin{equation}
T_{\mu \nu} (q,v) = \int d^4 x e^{ix{\cal Q}}
T \left[ (\bar{h}_v(x) \gamma_\mu (1-\gamma_5) q(x))
         (\bar{q}(0) \gamma_\mu (1-\gamma_5) h_v(0)) \right]
\end{equation}
where ${\cal Q} = q-m_Q v$ is the momentum transfer variable defining
the large momentum. 
We shall define the operator product expansion for the projection of
$T_{\mu \nu}$ needed to obtain the inclusive semileptonic rate
\begin{equation}
T (k,k^\prime,v) = T_{\mu \nu} (k+k^\prime,v) \Lambda^{\mu \nu}
\end{equation}
Before we write down the general form of the short distance expansion
in the effective theory we shall recall some basic formulae concerning
the spin structure in the effective theory. The sixteen covariants
built from Dirac matrices reduce to only four when projected into a
definite velocity sector. We shall use as a basis the unit matrix and
the three spin vectors $s_\mu$ with $v\cdot s =0$ which correspond 
in the rest frame $v = (1,0,0,0)$ to
the three Pauli matrices. The
translation table of the sixteen invarants is given by
\begin{eqnarray*}
1 \longrightarrow P_+ = \frac{1}{2} (1+\slash{v})
&\qquad &
\gamma_\mu \longrightarrow P_+ \gamma_\mu P_+ = v_\mu P_+
\\
\gamma_\mu \gamma_5 \longrightarrow P_+ \gamma_\mu \gamma_5 P_+ = s_\mu
&\qquad &
\gamma_5 \longrightarrow P_+ \gamma_5 P_+ = 0
\\
\sigma_{\mu \nu} \longrightarrow P_+ \sigma_{\mu \nu}P_+ =
       v^\alpha \epsilon_{\alpha \mu \nu \beta} s^\beta
\end{eqnarray*}
The spin vectors $s_\mu$ satisfy the relation
\begin{equation}
s_\mu s_\nu = (-g_{\mu \nu} + v_\mu v_\nu ) 1
              + i \epsilon_{\alpha \mu \nu \beta} v^\alpha s^\beta
\end{equation}
which is a generalization of the well known relation between the
Pauli matrices and which may be used to reduce products of spin
vectors.

At tree level one may obtain the operator product expansion up to
operators of dimension five by
considering the two diagrams gepicted in fig.1.
The result is
\begin{eqnarray} \label{ope}
T (k,k^\prime,v) &=&
\left( \frac{1}{{\cal Q}^2 - m_l^2 + i\epsilon} \right)
       64 (k^\prime {\cal Q})
              \bar{h}_v  \slash{k} (1-\gamma_5) h_v  \\
&+& \left( \frac{1}{{\cal Q}^2 - m_l^2 + i\epsilon} \right)
          64
              \bar{h}_v  (ik^\prime D) \slash{k} (1-\gamma_5) h_v
\nonumber \\
&-& \left( \frac{1}{{\cal Q}^2 - m_l^2 + i\epsilon} \right)^2
         128 (k^\prime {\cal Q})
              \bar{h}_v  (i{\cal Q}D) \slash{k} (1-\gamma_5) h_v
\nonumber \\
&-& \left( \frac{1}{{\cal Q}^2 - m_l^2 + i\epsilon} \right)^2
          64 (k^\prime {\cal Q})
              \bar{h}_v  (iD)^2 \slash{k} (1-\gamma_5) h_v
\nonumber \\
&-& \left( \frac{1}{{\cal Q}^2 - m_l^2 + i\epsilon} \right)^2
          64
  \bar{h}_v [(i{\cal Q}D) , (ik^\prime D)] \slash{k} (1-\gamma_5) h_v
\nonumber \\
&-& \left( \frac{1}{{\cal Q}^2 - m_l^2 + i\epsilon} \right)^2
    64 i \epsilon_{\lambda \alpha \beta \rho} {\cal Q}^\lambda
         k^{\prime \rho}
  \bar{h}_v (iD^\alpha) (iD^\beta) \slash{k} (1-\gamma_5) h_v
\nonumber \\
&-& \left( \frac{1}{{\cal Q}^2 - m_l^2 + i\epsilon} \right)^2
          128
  \bar{h}_v (ik^\prime D) (i{\cal Q} D) \slash{k} (1-\gamma_5) h_v
\nonumber \\
&+& \left( \frac{1}{{\cal Q}^2 - m_l^2 + i\epsilon} \right)^3
          256 (k^\prime {\cal Q} )
  \bar{h}_v (i{\cal Q} D)^2 \slash{k} (1-\gamma_5) h_v
\nonumber \\
&+& \mbox{Operators of Dimension six or higher} \nonumber
\end{eqnarray}
We note that the spin structure of all the terms is of the form
\begin{equation}
\bar{h}_v \gamma_\mu (1-\gamma_5) h_v =
v_\mu \bar{h}_v h_v -
\bar{h}_v s_\mu h_v
\end{equation}
reflecting the left handedness of the current.

The differential rate may now be calculated from the imaginary part
of the forward matrix element of the operator $T$
\begin{equation} \label{lepspec}
d \Gamma = \frac{G_F^2}{4 m_H} | V_{Qq} |^2
(2 \mbox{ Im } < H (v) | T (k,k^\prime , v) | H (v) >) d(PS)
\end{equation}
The relevant hadronic matrix elements will be discussed in the
next section.

\section{Hadronic Matrix Elements and Short Distance QCD Corrections}
We shall first consider the decay of a pseudoscalar meson. The only
nonvanishing matrix element of the dimension three operators is
\begin{equation} \label{opeme0}
<H(v) | \bar{h}_v h_v | H(v) > = 2 m_H .
\end{equation}
which is normalized as given above due to heavy quark symmetries.
Inserting this into (\ref{lepspec}) reproduces the parton model result.

All the matrix elements of the dimension four operators vanish.
The general form of the matrix element not involving the spin vector is
\begin{equation}
<H(v) | \bar{h}_v D^\alpha h_v | H(v) > = B_0 v^\alpha
\end{equation}
and by contraction with $v$ and using the equations of motion we have
$B_0=0$. The corresponding matrix element involving the spin vector
vanishes due to parity.

The matrix elements of the dimension five operators may be expressed
in terms of two independent parameters. Due to the equations of motion
we have
\begin{equation}
<H(v) | \bar{h}_v (iD^\alpha)(iD^\beta) h_v | H(v) > =
-2 m_H B_1 (g^{\alpha \beta} - v^\alpha v^\beta)
\end{equation}
where $B_1$\footnote{%
      This parameter is - up to a factor 3 - the parameter $\lambda_1$
      as defined in the paper by Falk and Neubert \cite{FN92}}
is determined by the trace
\begin{equation}
<H(v) | \bar{h}_v (iD)^2 h_v | H(v) > = - 6 m_H B_1 .
\end{equation}
This trace is related to the expectation value of the residual momentum
squared $k^2$ of the heavy quark in the heavy meson
($<k^2> \sim 3 B_1$) which is expected
to be of the order of $\Lambda_{QCD}^2$. Unfortunately one may not
extract a precise value from experimental data and one has
to rely on theoretical estimates. An estimate based on QCD sum
rules \cite{QCDsr} yields reltively large
value of $B_1 \sim 0.3$ GeV${}^2$.
We shall use a default value of
$B_1 \sim 0.2$ GeV${}^2$ and discuss the dependence of the result
on this parameter.

The other independent matrix elelement is parametrized as
\begin{equation}
<H(v) | \bar{h}_v (iD^\alpha)(iD^\beta) s^\rho h_v | H(v) > =
2 m_H B_2 \epsilon^{\lambda \alpha \beta \rho} v_\lambda
\end{equation}
where the constant $B_2$ is given in terms of the scalar matrix element
\begin{equation}
i \epsilon^{\lambda \alpha \beta \rho} v_\lambda
<H(v) | \bar{h}_v (iD_\alpha)(iD_\beta) s_\rho h_v | H(v) >
=  ig <H(v) | \bar{h}_v \sigma_{\mu \nu} G^{\mu \nu} h_v | H(v) >
\end{equation}
where $g$ is the strong coupling constant. 
This matrix element is related to the mass splitting between the
pseudoscalar meson and its spin symmetry partner vector meson
\begin{equation}
 ig <H(v) | \bar{h}_v \sigma_{\mu \nu} G^{\mu \nu} h_v | H(v) >
= -12 m_H (m_{B^*}^2 - m_B^2)
\end{equation}
from which we obtain
\begin{equation}
B_2 = \frac{1}{8} (m_{H^*}^2 - m_H^2 ) \sim 0.07 \mbox{ GeV}^2
\end{equation}
Thus lowest order corrections in the mesonic sector are given in
terms of two parameters, one of which is related to an
observable quantity.

The decay of a heavy ground state baryon $\Lambda_b$ has been considered
in \cite{MW93} in the approach proposed in \cite{BUV}.
For this decay
we have two matrix elements of the dimension three operators of which
the normalization is known due to heavy quark symmetries.
We have
\begin{eqnarray}
< \Lambda_H (v,s) | \bar{h}_v h_v | \Lambda_H (v,s) > &=& 2 m_H \\
< \Lambda_H (v,s) | \bar{h}_v s_\mu h_v | \Lambda_H (v,s) > &=&
2 m_H s_\mu
\end{eqnarray}
where $s_\mu$ is the direction of the spin of the $\Lambda_H$ baryon
in its rest frame.

As in the case of the mesons the matrix elements of the dimension
four operators vanish due to parity and the equations of motion.

Finally, the matrix elements of the dimension five operators are
given in terms of only one parameter
\begin{eqnarray}
< \Lambda_H (v,s) | \bar{h}_v (iD_\alpha) (iD_\beta) h_v
| \Lambda (v,s) > &=& - 2 m_H A_1 (g_{\alpha \beta}- v_\alpha v_\beta) \\
< \Lambda_H (v,s) | \bar{h}_v (iD_\alpha) (iD_\beta) s_\mu h_v
| \Lambda (v,s) > &=& - 2 m_H A_1 s_\mu (g_{\alpha \beta}- v_\alpha v_\beta)
\end{eqnarray}
since the heavy quark spin symmetry relates these two matrix elements.

The parameters $A_1$, $B_1$ and $B_2$ parametrize the long distance
QCD contributions to the inclusive semileptonic decay rates. Since
we are dealing with an operator product expansion the short distance
QCD contributions may be calculated as the scale dependence of
the Wilson coefficients, the tree level value of which may be
read off from (\ref{ope}). 

Short distance corrections may be considered using the usual machinery 
of the renormalization group. 
The calculation of the short distance QCD corrections may be found in
the literature. The anomalous
dimensions of the scalar operators defining the constants $A_1$, $B_1$
and $B_2$ has been performed
in \cite{FGL91}.

Due to reparametrization invariance \cite{repara} the operator $O_1$
\begin{equation}
O_1 = \bar{h}_v (iD)^2 h_v
\end{equation}
has vanishing anomalous dimension and thus $B_1$ as well as $A_1$
are scale invariant quantities.
The second operator $O_2$
\begin{equation}
O_2 = ig \bar{h}_v \sigma_{\mu \nu} G^{\mu \nu} h_v
\end{equation}
has nonvanishing anomalous dimension but does not mix with any
other operator. From the renormalization group equation with the
anomalous dimension from \cite{FGL91} 
\begin{equation} \label{RGE}
\mu^2 \frac{\partial}{\partial \mu^2}
B_2(\mu) = - \frac{3}{4 \pi} \alpha_s (\mu) B_2 (\mu)
\end{equation}
one obtains 
\begin{equation}
B_2 (\mu) = \left(\frac{\alpha_s (\mu)}{\alpha_s (m_H)}
         \right)^{9/(33-2n_f)} B_2 (m_H) .
\end{equation}
with $n_f = 4$ for $B$ hadron decays.
This scale dependence introduces an ${\cal Q}^2$ dependence, since
this is the relevant momentum transfer variable in the present case.
The effect of these short distance corrections will be discussed 
for the example of the lepton energy spectrum in
inclusive semileptonic decays of heavy mesons.

\section{An Application: Energy Spectrum in Heavy Meson Decays}
As an obvious application we shall consider the lepton energy spectrum
of the charged lepton
in an inclusive semileptonic decay of heavy mesons. To be specific we
shall consider the decay
$
\overline{B}^0 \to X_c e^- \bar{\nu}_e
$
Here $X$ is a hadronic final state containing charm.

The differential
decay rate is given by
\begin{equation} \label{diffrate}
\frac{d \Gamma}{dE}  = \frac{G_F^2}{4 m_H} | V_{Qq} |^2
\int \widetilde{dk} \widetilde{dk}^\prime \delta({\cal Q}^2 - m_l^2)
\delta (E-vk^\prime)
(2 \mbox{ Im } < H (v) | T (k,k^\prime , v) | H (v) >)
\end{equation}

The calculational details of the phase space integration are not 
very iluminating and are given in an appendix. 
However, unlike in the usual case of phase space integrals we
have due to the relations
\begin{eqnarray}
2 i \mbox{Im} \left(\frac{1}{{\cal Q}^2 - m_l^2 + i\epsilon} \right) 
&=& - (2 \pi) \delta^\prime ({\cal Q}^2 - m_l^2 )  \\
2 i \mbox{Im} \left(\frac{1}{{\cal Q}^2 - m_l^2 + i\epsilon}\right)^2 
&=& (2 \pi) \delta^\prime ({\cal Q}^2 - m_l^2 ) \\
2 i \mbox{Im} \left(\frac{1}{{\cal Q}^2 - m_l^2 + i\epsilon}\right)^3 
&=& - \pi \delta^{\prime \prime} ({\cal Q}^2 - m_l^2 )
\end{eqnarray}
also integrations involving the first and second deriavtive of the 
delta function $\delta^\prime$ and $\delta^{\prime \prime}$. Flipping
the derivative yields derivatives of the integrands. Since the QCD
short distance corrections imply a dependence of $B_2$ on the kinematic
variables one has to differentiate $B_2$ as well. This yields an 
additive correction term which is proportional to $\alpha_s$ taken 
at the scale ${\cal Q}^2$; this scale is fixed to ${\cal Q}^2 = m_l^2$
due to the delta functions showing that the present approach is 
indeed limited to $m_l \gg \Lambda_{QCD}$.  

The results for the lepton spectrum is displayed in fig.2 and 3.
Fig.2
shows the charged lepton energy spectrum for the inclusive semileptonic decay
of a $\overline{B}^0$ or a $B^-$ into charmed final states. The
parameters used are $m_B = m_b = 5.28$ GeV, $m_l = m_c = m_D = 1.86$
GeV, $|V_{cb}| = 0.043$ and $B_1 = 0.2$ GeV${}^2$. The solid line (a)
is the full result, including long- and short distance corrections.
This is compared to the spectrum including the corrections only
at tree level, i.e.~without the QCD short distance contribution
(line (b)). Finally, curve (c) is the leading order result which is
the spectrum obtained from the parton model. The spectrum obtained
this way is slightly shifted to lower energies and the short distance
corrections yield a small enhancement at low energies. Qualitatively
this result agrees with the one obtained in \cite{BUV}.

%\begin{figure}
%\vskip 8cm
%\caption{The lepton energy spectrum of the inclusive $b \to c$ semileptonic
%         decay. The parameters used are $m_b = m_B = 5.28$ GeV,
%         $m_c = m_D = 1.86$ GeV, $|V_{cb}| = 0.043$
%         and $B_1 = 0.2$ GeV ${}^2$. The curves are:
%         (a) Full result including long- and short distance
%         contributions,
%         (b) Result without short distance QCD corrections,
%         (c) leading order result, i.e.~ the parton model result.}
%\label{fig2}
%\end{figure}

Finally, we shall investigate the dependence on the parameter $B_1$
by considering the total rates. In fig.3 we plot the quantity
$$
\frac{\Gamma_{corr} (\overline{B}^0 \to X_c e^- \bar{\nu}_e)}
     {\Gamma_{parton} (\overline{B}^0 \to X_c e^- \bar{\nu}_e)} - 1
$$
in percent as a function of the value of $B_1$.
We chose the same parameters as for fig.2, except that 
$B_1$ is varied in the range $0.04 \le B_2 \le 0.4$.
Curve (a) is the full result, including long- as well as short 
distance contributions, curve (b) conbtains the tree level corrections
only. The corrections reduce the total rate (compared
to the parton model rate) for values of $B_1$ above 0.1. This seems to 
favor large values of $B_1$, since a reduction of the semileptonic 
branching fraction is phenomenologically welcome. Furthermore, one 
could in turn try to obtain matrix element corresponding to $B_1$ 
from a preceise measurement of the inclusive lepton spectrum.

%\begin{figure}
%\vskip 8cm
%\caption{The ratio of the corrected total decay rate and the parton
%         model rate for the inclusive decays into charmed and
%         noncharmed final states as a function of the parameter $B_2$.
%         The curves correspond to:
%         (a) $B^- \to X_c e^- \bar{\nu}_e$, full result,
%         (b) $B^- \to X_c e^- \bar{\nu}_e$, corrections only at tree
%             level.}
%\label{fig4}
%\end{figure}

\section{Conclusions}
Inclusive semileptonic decays
of heavy hadrons may be described by employing
an operator product expansion for the operators appearing in the
hadronic tensor. The expansion is performed after switching to
heavy quark effective field theory where the large momentum of the
heavy hadron is scaled out. The relevant momentum transfer variable
is ${\cal Q} = q - m_Q v$ where $q$ is the momentum transferred to
the leptons.

In the present paper the region of phase space has been considered
in which ${\cal Q}^2 \ll m_Q^2$ and heavy quark effective field theory
to leading order in the $1/m_Q$ expansion is sufficient.
The approach of \cite{BUV} and \cite{BS93} is in fact valid in a different portion
of phase space, since \cite{BUV,BS93} assume that
${\cal Q}^2 \sim m_Q^2$ and thus a simultaneous expansion in
$1/m_Q$ and $1/{\cal Q}^2$ is necessary.

The lepton spectrum as obtained in \cite{BUV,BS93} contains delta functions
and its derivatives which reflects the failure of the expansion at
the edge of phase space $E \sim E_{max}$. In contrast to that the
approach applied here yields a completely smooth lepton energy
spectrum without any delta function singularities.

Of course, the limitations of the procedure proposed here
(namely ${\cal Q}^2 \ll m_Q^2$) do show up at some point.
If one studies for example the rate differential in
the momentum transfer $q^2 = (k+k^\prime)^2$ and the total leptonic
energy $vq = vk + vk^\prime$ then one finds also delta functions and
its derivatives entering the differential rate using the method chosen
here. However, after an integration over the total leptonic energy one
obtains a smooth $q^2$ spectrum and the only reminder for the
breakdown of the ${\cal Q}^2$ expansion is a behaviour of the correction 
term of the type
\begin{equation}
\frac{d \Gamma}{dq^2 } \sim \frac{1}{\sqrt{(m_Q - m_l)^2 - q^2}}
\end{equation}
at the endpoint of the leptonic $q^2$  spectrum.
This is an integrable singularity and hence a finite total rate may
be obtained as expected.

In conclusion, depending on the region of phase space considered
one has to choose between the method proposed in \cite{BUV,BS93} and
the one presented here. To obtain the lepton spectrum one has to
integrate over the phase space and 
phase space regions contribute where the approximations used become
invalid. However, both methods yield finite total rates which means
that the singularities in the lepton spectrum are intagrable and
one may still expect a reasonable lepton energy spectrum from both
methods; this expectation is confirmed by the qualitative agreement
of the lepton energy spectra obtained from both approaches.

\section*{Acknowledgements}
I like to thank M. Wise for clarifying discussions and for informing
me about ref.~\cite{MW93} prior to publication. I am also indebted to 
B. Bloch for pointing out ref.\cite{BS93} to me.  
I also like to thank W. Kilian and P. Manakos for valuable comments.

\appendix

\section{Some Details of the Calculation of the Lepton Energy
         Spectrum}
In this appendix we give some of the technical details of the 
calculation for the energy spectrum of the charged lepton.
 
In order to perform the phase space integrations in  (\ref{diffrate}) 
it is convenient to define three integrals of some function $F$ 
of the three independent scalar products $vk$, $vk^\prime$ and 
$2kk^\prime$
$$
F = F(vk,vk^\prime,2kk^\prime).
$$
We define the integral
\begin{eqnarray}
I_0 [F] &=&  \int \widetilde{dk} \widetilde{dk}^\prime
\delta({\cal Q}^2 - m_l^2)
\delta (E-vk^\prime)
F(vk,vk^\prime,2kk^\prime) \\ \nonumber
        &=& \frac{1}{32 \pi^3} \int\limits_\alpha^\beta dk
          F(k,E,2m_H (E-E_{max}+k^\prime))
\end{eqnarray}
where the limits of the integration are given by
\begin{equation}
\alpha = m_H \frac{E_{max} - E}{m_H - 2 E} \qquad
\beta = E_{max} - E
\end{equation}
By definition
$E$ is the energy of the charged lepton in the rest frame of the
decaying heavy meson and $E_{max}$ is its maximal energy
\begin{equation}
E_{max} = \frac{1}{2m_H} (m_H^2 - m_l^2)
\end{equation}
The corrections contain derivatives of delta functions and it is
convenient to define two more integrals
\begin{eqnarray}
I_1 [F] &=&  \int \widetilde{dk} \widetilde{dk}^\prime
\delta^\prime ({\cal Q}^2 - m_l^2)
\delta (E-vk^\prime)
F(vk,vk^\prime,2kk^\prime) \\
I_2 [F] &=&  \int \widetilde{dk} \widetilde{dk}^\prime
\delta^{\prime \prime} ({\cal Q}^2 - m_l^2)
\delta (E-vk^\prime)
F(vk,vk^\prime,2kk^\prime)
\end{eqnarray}
where $\delta^\prime$ and $\delta^{\prime \prime}$ denotes the first
and second derivative of the delta function. These two integrals
may be expressed in terms of the integral $I_0$
\begin{eqnarray} \label{I1}
I_1 [F] &=& - \frac{1}{64 \pi^3} \left( \frac{F(\beta,E,0)}{m_H}
            - \frac{F(\alpha,E,4 E \alpha)}{m_H - 2E} \right)
            - I_0 [F^\prime] \\
I_2 [F] &=&  \frac{E}{32 \pi^3} \delta (2m_H (E_{max}-E)) F(0,E,0)
             \frac{1}{m_H (m_H - 2E)}
            + I_0 [F^{\prime \prime}]
\end{eqnarray}
where $F^\prime$ and $F^{\prime \prime}$ denotes the first and second
derivative of $F$ with respect to the last argument $2kk^\prime$.

Inserting the operator product expansion (\ref{ope}) one finds
that the piece containing the dimension three operators reproduces
the parton model result
\begin{equation}
\frac{d \Gamma}{dE}  = - 32 G_F^2 | V_{Qq} |^2
                       I_0 [ (vk)(kk^\prime - m_H vk^\prime) ]
\end{equation}

At tree level the correction terms originating from the dimension
five operators are given by
\begin{eqnarray}
\Delta \frac{d \Gamma}{dE}  &=&  32 G_F^2 | V_{Qq} |^2 \left\{ B_1
I_1 [ 5(vk)(k^\prime {\cal Q}) - 2 (vk)(vk^\prime)(v {\cal Q}) ] \right.
\\ \nonumber &+& B_2 \left.
I_1 [ 2(vk^\prime)(k {\cal Q}) - 2 (kk^\prime)(v {\cal Q})  ] \right\} \\
\nonumber
&-&  64 G_F^2 | V_{Qq} |^2 B_1
I_2 [ (k^\prime {\cal Q})(vk)((v {\cal Q})^2 - {\cal Q}^2) ]
\end{eqnarray}

The QCD short distance contributions may be included by
putting in the scale dependence of $B_2$ as discussed in the last
section. At the one loop level one finds that
aside from the fact that
the value of $B_2$ is changed additional terms originating
from the derivative in (\ref{I1}) acting on $B_2$ occur, since
now $B_2$ depends on $2kk^\prime$.
The derivative of $B_2$ is obtained from the renormalization group
equation (\ref{RGE}).
Inserting this we obtain an additional term due to the short
distance contribution to $B_2$
\begin{equation}
\Delta_{SD} \frac{d \Gamma}{dE}  =  - 32 G_F^2 | V_{Qq} |^2
I_0 [ 2(vk^\prime)(k {\cal Q}) - 2 (kk^\prime)(v {\cal Q})  ]
\frac{3}{4 \pi m_l^2} \alpha_s (m_l^2) B_2 (m_l^2)
\end{equation}
Here and in the other terms where it occurs
$B_2$ has to be scaled down to ${\cal Q}^2$ which is due to
the delta functions
of phase space fixed to the value $m_l^2$.

\newpage

\section*{Figure Captions}
\begin{itemize}
%\begin{figure}
%\vskip 8cm
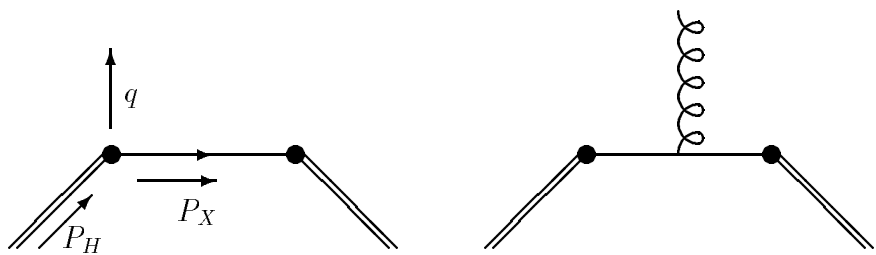
\item[Fig.1] {Feynman Diagrams to be evaluated to obtain the Wilson
         coefficients of the operators up to dimension five. The blob 
         denotes the left handed current and the double line denotes 
         the decaying heavy quark.}
%\label{fig1}
%\end{figure}

%\begin{figure}
%\vskip 8cm
\item[Fig.2]{The lepton energy spectrum $ d\Gamma / dE \times 10^{15}$ 
         of the inclusive semileptonic
         decay $\overline{B}^0 \to X_c e^- \bar{\nu}_e$. 
         The parameters used are $m_b = m_B = 5.28$ GeV,
         $m_c = m_D = 1.86$ GeV, $|V_{cb}| = 0.043$
         and $B_1 = 0.2$ GeV ${}^2$. The curves are:
         (a) Full result including long- and short distance
         contributions,
         (b) Result without short distance QCD corrections,
         (c) leading order result, i.e.~ the parton model result.}
%\label{fig2}
%\end{figure}

%\begin{figure}
%\vskip 8cm
\item[Fig.3]{The ratio of total rates 
             $(\Gamma_{corr} / \Gamma_{Parton}) - 1$ in percent for the 
             inclusive decay $\overline{B}^0 \to X_c e^- \bar{\nu}_e$
             function of the parameter $B_1$.
         The curves correspond to:
         (a) full result,
         (b) corrections only at tree
             level. }
%\label{fig4}
%\end{figure}
\end{itemize}
\end{document}